\documentclass{article}
\usepackage{amsfonts}
\usepackage{epsfig}

\title{A Weak Quantum Oblivious Transfer}
\author{K.Y. Cheong 
\footnote{School of Information Science, Japan Advanced Institute of Science and Technology. Email: kaiyuen@jaist.ac.jp}
\and Min-Hsiu Hsieh 
\footnote{University of Cambridge. Email: minhsiuh@gmail.com} 
\and Takeshi Koshiba 
\footnote{Division of Mathematics, Electronics and Informatics, Graduate School of Science and Engineering, Saitama University, Japan. 
Email: koshiba@mail.saitama-u.ac.jp}}

\begin{document}
\maketitle
\begin{abstract}
Due to the commonly known impossibility results, information theoretic security is considered impossible for oblivious transfer (OT)
in both the classical and the quantum world. In this paper, we proposed a weak version of the all-or-nothing OT.
In our protocol the honest parties do not need long term quantum memory, entanglements, or sophisticated quantum computations.
We observe some difference between the classical and quantum OT impossibilities.

\noindent
{\bf Keywords:} quantum oblivious transfer, information theoretic security
\end{abstract}

\section{Introduction}
Oblivious Transfer (OT) is an important two-party cryptographic protocol, as a building block for many general cryptographic primitives. 
In the first OT system introduced by Rabin \cite{rabin}, a message is received with probability 1/2 and the sender
does not know whether the message reaches the receiver. This is later called the all-or-nothing OT or simply the Rabin OT.
Even et al. \cite{x} defined the 1-out-of-2 OT, where the sender has two secrets and the receiver can choose one and only one of them
in an oblivious manner. That is, the sender cannot know the receiver's choice 
and the receiver cannot know anything more than one secret. The two types of OT are shown to be equivalent in the classical world \cite{eq}, 
in the sense that one form of OT can be used as a building block to construct the other.

For both types of OT in the classical world, it is rather obvious that unconditional, information theoretic security cannot be achieved for both sides at the same time. 
Therefore, computational assumptions such as the existence of trapdoor function is required. In the practical sense, the hardness of factorization
and discrete logarithm are popular assumptions that are often required in many cryptographic protocols.
Such assumptions are deeply threatened by the development of quantum computing, due to the Shor's algorithm \cite{shor}.

On the other hand, quantum techniques also provide new potential tools for the construction of cryptographic primitives. 
For example, the BB84 protocol \cite{bb84} is proposed for key agreement with unconditional security, which is impossible in the classical world. 
Therefore, there is hope that quantum cryptography will develop faster than quantum cryptanalysis, and 
a new kind of cryptography will be ready to replace the old one before it turns out to be insecure.
Moreover, it is interesting to note that the original idea of OT comes from the quantum realm, in the novel paper by Wiesner \cite{wiesner}.

\subsection{The background story of quantum OT}
Based on the BB84 key agreement, in 1994 Cr\'{e}peau \cite{c94ot} proposed a novel protocol for 1-out-of-2 OT which resolves most problems
known in previous schemes. The security of this scheme relies on the use of another important cryptographic primitive, the Bit Commitment (BC).
Since then, in \cite{sm} and \cite{yao} it has been formally proved that \cite{c94ot} is secure with the use of quantum BC.
The construction of OT from BC using quantum techniques is itself an achievement, since such a construction is not known in the classical world.

Unfortunately, in 1997 a few major impossibility results were found concerning OT and BC. These include the work of Lo against the 
1-out-of-2 OT \cite{lo} and the work of Mayers \cite{mayers} against BC. Lo and Chau \cite{lo2} also independently
argued that unconditionally secure quantum BC is impossible. This is a major setback and breakthrough in the research of 
quantum based OT, BC, and general cryptographic protocols. 

Since then, new protocols have been proposed, avoiding the known impossibility results in various ways.
For example, quantum BC can be secure based on computational complexity assumptions \cite{owp} or physical assumptions about the 
technology used by the adversary \cite{phy}. 

On the other hand, currently there is no known impossibility results against the Rabin OT. It is unknown if the equivalence between 
Rabin OT and 1-out-of-2 OT in the classical world applies to the quantum world. If this is the case, then quantum Rabin OT is also impossible.
In this paper, to investigate the possibility of Rabin OT, we study a weakened form of Rabin OT.

\subsection{Our contribution}
In \cite{c94ot}, the idea of the BB84 protocol is used to construct OT. The attack of delayed measurement is the major reason 
that BC is required. Without BC, there is no suitable time for sending the information about the basis of the qubits.
In this paper, we propose a weak Rabin OT protocol based on the B92 key agreement protocol \cite{b92}. Our observation is that,
unlike the BB84 protocol, in the B92 scheme the sender never needs to send the information about the basis of the qubits.

Despite the general impossibility, there are some known differences for OT in the classical and quantum world. 
In the classical world, OT is impossible even in the honest-but-curious model, where the parties must follow the protocol, 
but could try to gain more information through private computations. In the quantum world, if the parties follow the protocol strictly,
OT would be possible.

In this paper, we show that our weak OT protocol has some properties that are not possible in the classical world. 
This is another difference between the classical and quantum OT impossibilities. It also suggests that an impossibility result for Rabin OT
may not be easy to obtain, since a weak version of it is actually possible.

\section{The weak Rabin OT protocol}
\subsection{Definition}
There may be several ways to weaken the security definition of Rabin OT. For the sake of our study, we give the following definition
for the weak Rabin OT (WROT) protocol, between sender Alice and receiver Bob.
\begin{enumerate}
\item
The honest Alice inputs a random message bit $x$.
\item
The honest Bob receives $x$ correctly from the honest Alice with probability $p$, a value specified in the protocol. 
Otherwise Bob receives zero information about $x$. Bob knows if he has received $x$. If he does not receive $x$, he outputs $\bot$.
\item
For any run of the protocol, a cheating Alice has limited advantage $v$ to guess or change the probability that Bob outputs $\bot$. 
That is, in Alice's final view, if the probability that Bob outputs $\bot$ is $p'$, then $|p'-(1-p)| \leq |v|$. 
Note that a cheating Alice needs not prepare any message bit.
\item
A cheating Bob always tries to increases the overall rate that he gets $x$, possibly through guessing. 
He needs not be certain about the correctness of the output, so he never outputs $\bot$.
For any cheating Bob, his advantage is limited to $u$. If the final rate that he gets $x$ correctly is $q$, then $q \leq p + \frac{(1-p)}{2}+u$.
\end{enumerate}

Under this definition, Alice can cheat either by trying to find out what happens at Bob's side when he decodes the bit, 
or by really changing the probability that Bob gets $\bot$. In the WROT, Alice can change the probability that Bob gets $\bot$ by $v$. 
Also, Bob can increase the rate he gets $x$ to $q > p+\frac{1-p}{2}$. This is why the protocol is called weak.

\subsection{The WROT construction}
At the beginning the two parties specify two non-orthogonal quantum states $|\psi_0\rangle$ and $|\psi_1\rangle$ of one qubit.
They use $|\psi_0\rangle$ to represent bit 0 and $|\psi_1\rangle$ to represent 1.
According to $x$, Alice prepares $|\psi_0\rangle$ or $|\psi_1\rangle$ to send to Bob, and 
Bob uses the Positive Operator-Valued Measure (POVM) method to distinguish the two states \cite{povm} unambiguously.
Since only the angle $\alpha < \frac{\pi}{2}$ between vectors $|\psi_0\rangle$ and $|\psi_1\rangle$ affects Bob's ability to distinguish them, we set 
$|\psi_0\rangle=|0\rangle$ and $|\psi_1\rangle=a|0\rangle+\sqrt{1-a^2}|1\rangle$, with $a=\cos \alpha$.
In this case the POVM elements are:
\begin{eqnarray}
E_1&=&\frac{1}{1+a}|1\rangle \langle1| \cr
E_2&=&\frac{1}{1+a}(\sqrt{1-a^2}|0\rangle-a|1\rangle)(\sqrt{1-a^2} \langle 0| - a \langle 1|) \cr
E_3&=&I-E_1-E_2
\end{eqnarray}
where the decode probability for Bob is $p=1-a$ in the honest case. In other words, Bob gets $\bot$ with probability $a$. 
The measurement matrices $(E_1,E_2,E_3)$ represent the output symbols $(1,0,\bot)$ of Bob, respectively. The actual implementation of the 
POVM is not important. The method in \cite{povm} suggests that Bob only needs a unitary operation of two qubits followed by two measurements
of one qubit.

\subsection{Security properties}
First, we argue that Alice cannot use entanglement for cheating.
Notice that in our simple scheme, Alice sends a qubit to the honest Bob and he measures it with the POVM. If the qubit is entangled to 
some quantum states held by Alice, the cheating Alice should perform her measurement on such quantum states after Bob finishes his. But due to fundamental physical laws,
Alice could never receive any information about whether Bob has performed a measurement. So there is no difference if Alice measures before Bob does.
In that case, Alice would have created and collapsed the entanglement at the same time, before the qubit is sent out.
This gives her no use of the entanglement. This argument is valid for any implementation of the POVM at Bob's side.

Instead of using entanglement, Alice would better create a pure state qubit $|\psi \rangle$ she wants to send to Bob.
For this case, $|\psi \rangle$ can be any one qubit state. In general $|\psi \rangle$ would be 
$\sqrt{1-d_1^2-d_2^2}|0\rangle + (d_1+id_2)|1\rangle$ where $d_1,d_2$ are real numbers and $d_1^2+d_2^2 \leq 1$. 
The probability of Bob getting $\bot$ is changed to $\langle \psi|E_3|\psi \rangle$.

We can compute Alice's advantage $v$ by the difference between this probability and the value $a$, 
the rate of Bob getting $\bot$ in the honest case. A direct calculation gives 
\begin{equation}
v=\langle \psi|E_3|\psi \rangle-a=\frac{2a(-ad_1^2-ad_2^2+d_1\sqrt{(1-a^2)(1-d_1^2-d_2^2)})}{1+a}. \label{v}
\end{equation}

If Alice is cheating, she could either increase or decrease the chance that Bob gets $\bot$. 
She can always make sure that Bob does not get $\bot$ at all, when $\langle \psi |E_3| \psi \rangle=0$.
The lowest (negative) value of $v$ is always $-a$. From (\ref{v}), the highest value of $v$, given $a$, occurs when $d_1=\sqrt{\frac{1-a}{2}}$ and $d_2=0$.
In this case, $v=a (\frac{1-a}{1+a}) <a$. In the WROT, Alice can freely choose any $v$ between the maximum and minimum.
Since $-a \leq v < a$, Alice's advantage $|v|$ can be controlled by the choice of $a$ in the protocol. A smaller $a$ provides better security against Alice.

Next, we consider a cheating Bob, who would never output $\bot$, but would rather try to guess $x$. He accepts some uncertainty, which is
unavoidable since the two given states are non-orthogonal. To provide the lowest error rate guessing $x$, he chooses an optimized projection
with orthogonal basis $|\phi_0\rangle$ and $|\phi_1\rangle$. Optimization is achieved \cite{projection} when $\sum_i \sin^2{\theta_i}$ is minimized,
where $\theta_i$ is the angle between $|\psi_i \rangle$ and $|\phi_i \rangle$ for $i \in \{0,1\}$.
This happens when $\theta_0=\theta_1=\frac{\pi}{4}-\frac{\alpha}{2}$. Call this angle $\theta$, the error rate of Bob guessing $x$ is $\sin^2 \theta$. Therefore $q=\cos^2 \theta.$

Here, we observe a difference between classical and quantum world concerning the WROT. For the classical world, even in the
honest-but-curious model of weaker attacks, Alice can compute exactly what Bob can, based on the communication between them. 
That is, Alice knows Bob's view about the random variable $x$. Therefore, either Alice can know for certain that 
Bob outputs $\bot$, or Bob can find a way to compute $x$. In terms of the WROT, it is either $v=p$ or $q=1$. 

In the quantum case based on our scheme, Alice never has complete information of Bob's view on $x$, as the result of the POVM is unpredictable.
It can be seen that $v$ can be much smaller than $p=1-a$, while $q \neq 1$ as there is no way to perfectly distinguish
non-orthogonal quantum states. 

\begin{figure}
\epsfig{file=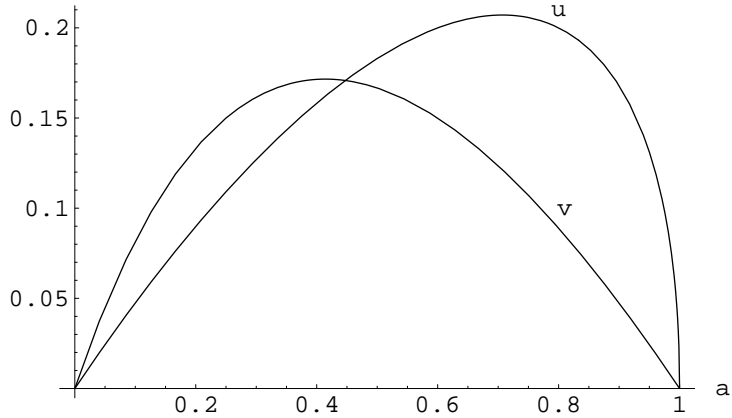,height=6cm}
\caption{Relation of $v$ and $u$ with respect to $a$.}
\label{fig:one}
\end{figure}

We can study the relation between $a$, the maximized $v$, and the maximized $u$. Figure \ref{fig:one} shows the graph of $u$ and $v$ against $a$. 
While a trade-off between $u$ and $v$ may be expected, in our scheme both $u$ and $v$ can be suppressed with a lower $a$. 
But this is not at all a good news because, in very vague terms, when $a$ is too small, the usefulness of the Rabin OT is also low, as the uncertainty about 
whether Bob receives $\bot$ is small in Alice's point of view. Note that the graph only plots the maximum value of $v$, not the minimum. 
The minimum value is $v=-a$ and it is significant when $a$ is larger. Therefore it is reasonable to consider only small $a$.

\section{Relation with OT impossibility}
If the WROT is used to construct 1-out-of-2 OT using standard techniques \cite{eq}, a weak 1-out-of-2 OT will be resulted,
but the weakness parameters of the obtained scheme would be rather high. Therefore it does not meet the conditions for enhancement
to normal OT \cite{wot}. In this way, our protocol neither violates the impossibility of 1-out-of-2 OT nor suggests inequivalence of Rabin OT and 1-out-of-2 OT.
On the other hand, it is currently unknown if the normal Rabin OT is possible or not. But to seek any impossibility result one must avoid our
constructive result of weak Rabin OT.

\end{document}